\documentclass[12pt,a4paper]{article}

\usepackage[utf8]{inputenc}
\usepackage[T1]{fontenc}
\usepackage{amsmath,amssymb}
\usepackage{graphicx}
\usepackage{booktabs}
\usepackage{hyperref}
\usepackage[margin=1in]{geometry}
\usepackage{natbib}
\usepackage{float}
\usepackage{caption}
\usepackage{setspace}
\usepackage{xcolor}

\hypersetup{
    colorlinks=true,
    linkcolor=blue,
    citecolor=blue,
    urlcolor=blue
}

\title{\textbf{Human genetic evidence is associated with drug approval across therapeutic areas: an observational analysis of 26,278 target-disease pairs with temporal validation and feature ablation}}

\author{Victoria Paterson\\
School of Informatics, University of Edinburgh, Edinburgh, UK\\
\texttt{v.paterson@sms.ed.ac.uk}}

\date{}

\begin{document}

\maketitle

\begin{abstract}
Genetic evidence is enriched among approved drug targets: in an observational analysis of 26,278 target--disease pairs from Open Targets and ChEMBL, targets with any genetic association had a 3.25-fold higher approval rate than those without (OR~$= 3.25$, 95\% CI 2.79--3.79, $p = 1.91 \times 10^{-42}$). A target-level analysis accounting for the non-independence of pairs sharing the same gene---26,278 pairs derive from only 1,434 underlying genes---gave OR~$= 2.79$ (bootstrap 95\% CI 2.22--3.53), confirming the association after correcting for repeated-gene contributions; the oncology pair-level OR of 6.72 attenuates to 2.71 at the target level, illustrating how non-independence inflates area-specific pair-level estimates. The enrichment replicated in an independent cohort of post-2015 approvals (OR~$= 3.51$, $p = 1.72 \times 10^{-8}$). Feature ablation across six evidence types revealed that literature mining alone accounts for most classifier performance (AUPRC~$= 0.099$ versus $0.109$ for all features combined), consistent with temporal leakage from post-approval publications; because Open Targets literature scores carry no timestamps, this leakage component cannot be fully isolated. Excluding literature, the remaining evidence types retain above-baseline signal (AUPRC~$= 0.084$, $1.63\times$ baseline). Sensitivity analyses on Phase~3 inclusion and negative-class restriction bracket the pair-level OR between 3.25 and 4.93. Genetic evidence alone yields only a 1.0-percentage-point absolute AUPRC gain, and the best-performing model has a Brier score of 0.20 with poor calibration; the classifier has limited practical predictive value. We catalogue 1,433 genetically supported Phase~1/2 pairs as a hypothesis-generating resource. All findings are observational.

\end{abstract}

\newpage

\section{Introduction}

Drugs targeting genes with human genetic support are approved at roughly twice the rate of those without, an observation first reported by Nelson et al.~\citep{nelson2015} and subsequently replicated~\citep{king2019,minikel2020}. This finding has shaped target-selection practice across the pharmaceutical industry, but several analytical gaps remain. The original studies treated genetic evidence as a binary variable and did not compare its contribution against other evidence types within a unified framework. Platforms such as Open Targets~\citep{ochoa2023} now integrate genetic associations alongside five additional evidence types---somatic mutations, literature mining, RNA expression, animal models, and affected pathways---yet the relative contribution of each source to the enrichment signal has not been systematically compared. A further concern is that literature mining scores aggregate publication counts without distinguishing pre-clinical from post-approval activity, potentially inflating apparent associations through temporal leakage. Finally, prior analyses have not addressed the non-independence introduced when many target--disease pairs share the same gene, which can distort both statistical tests and cross-validation estimates.

This study does not attempt to build a clinical prediction tool. Drug approval is determined by formulation, pharmacokinetics, commercial strategy, and regulatory dynamics that extend well beyond target biology, so modest classifier performance is both expected and unremarkable~\citep{nelson2015,king2019}. The goal is analytical: to use interpretable models as a lens for quantifying observational associations between evidence types and approval outcomes, and for isolating whatever signal remains once the literature-leakage component is removed. The analysis is cross-sectional and observational throughout; no causal claims about the role of genetic evidence in drug approval are warranted.

The study extends prior work in three ways. First, an ablation and temporal-validation framework provides the first systematic estimate of literature-leakage magnitude in this domain, comparing six evidence types within a single dataset. Second, a target-level sensitivity analysis addresses the non-independence of pairs sharing the same gene, a methodological issue not previously examined in this literature. Third, we test whether the genetic enrichment signal is consistent across five therapeutic areas using both pair-level and target-level analyses. We also release a catalogue of 1,433 genetically supported target--disease pairs at Phase~1/2 as a hypothesis-generating resource. The clinical attrition problem motivating this work is well established~\citep{hay2014,dimasi2016}; roughly 90\% of Phase~1 candidates fail, most for lack of efficacy.

We assemble 26,278 target--disease pairs across five therapeutic areas, label them with clinical-phase outcomes from ChEMBL, annotate them with Open Targets evidence scores, and analyse the dataset with gradient-boosted trees, logistic regression, SHAP interpretability, systematic feature ablation, and temporal validation.

\section{Methods}

\subsection{Data Sources}

\textbf{Open Targets Platform.} The Open Targets Platform (accessed April 2024 via the v4 GraphQL API; data release approximately corresponding to version 24.03)~\citep{ochoa2023} provided two types of records: drug--target--disease--phase mappings from the \texttt{drug} endpoint (\texttt{mechanismsOfAction} and \texttt{indications} fields), and multi-source evidence scores for target--disease associations from the \texttt{disease.associatedTargets} endpoint. Evidence scores are pre-computed aggregates in $[0,\,1]$ for six data types: genetic association, somatic mutation, literature mining, RNA expression, animal model, and affected pathway.

\textbf{ChEMBL.} Drug--indication pairs were retrieved from ChEMBL (v33)~\citep{mendez2019} using the \texttt{chembl-webresource-client} Python library. Molecule identifiers from the drug indication table were intersected with those from the drug mechanism table to retain compounds with both indication and target annotations.

\subsection{Dataset Construction}

For each molecule in the intersection of ChEMBL indication and mechanism tables ($n = 4{,}622$), the Open Targets drug endpoint was queried to obtain linked targets (Ensembl gene IDs), linked diseases (EFO/MONDO/HP identifiers), and maximum clinical stage per indication. Clinical stage strings were mapped to integer phases 1--4. For each unique (Ensembl ID, EFO ID) pair, the maximum phase achieved across all associated drugs was recorded. Pairs for which no evidence scores were available in the Open Targets association table were dropped; within retained pairs, missing individual evidence types were set to zero.

\textbf{Label definition.} A pair was labelled \emph{positive} if it reached Phase~4 (regulatory approval; $n = 1{,}364$) and \emph{negative} if it reached only Phase~1 or Phase~2 ($n = 24{,}914$). Phase~3 pairs ($n = 12{,}125$) were excluded from the primary analysis as outcome-ambiguous; a sensitivity analysis including them as negatives is reported in Section~\ref{sec:sensitivity}. The final dataset contains 26,278 pairs (5.2\% positive).

\textbf{Negative-class heterogeneity.} The negative class is not homogeneous. Phase~1/2 pairs include true clinical failures, commercially abandoned programmes, and projects still under active development. ChEMBL's coverage of early-phase failures is incomplete, so the composition of this class cannot be fully characterised. To assess the impact of this heterogeneity, we identified Phase~1/2 pairs where the associated drug had been approved for a different indication before 2015 but the pair itself had not advanced---a proxy for likely-terminated programmes. This subset comprised 16,693 of the 24,914 negatives (67.0\%). Because misclassifying ongoing programmes as failures pulls odds ratios toward the null, our enrichment estimates are probably conservative. This issue is revisited in Section~\ref{sec:discussion}.

\subsection{Therapeutic Area Assignment}

Disease identifiers were classified into five therapeutic areas (Oncology, CVRM, Respiratory, Rare Disease, and Immunology) using ontology-based annotation from the Open Targets Platform. Each disease's \texttt{therapeuticAreas} field was queried via the GraphQL API. Diseases not matching any mapped category were assigned ``Other.''

\subsection{Machine Learning Models}

Two classifiers were evaluated: logistic regression with standard scaling, balanced class weights, and a maximum of 1,000 iterations; and XGBoost~\citep{chen2016} with 300 estimators, maximum depth~4, learning rate~0.05, subsample fraction~0.8, column-sample-by-tree fraction~0.8, and \texttt{scale\_pos\_weight} set to the negative-to-positive class ratio (18.3). Both were assessed by five-fold cross-validation grouped by target gene (\texttt{StratifiedGroupKFold}), ensuring that no target gene appears in both training and test folds. This target-grouped design is a methodological strength of the study: the 26,278 pairs derive from only 1,434 unique target genes (mean 18.3 pairs per gene), so standard stratified CV would allow information about a gene in the training set to inform predictions for other pairs sharing that gene. For comparison, standard pair-level stratified CV results are also reported. All random operations used seed~42. The primary metric was the area under the precision--recall curve (AUPRC), which is more informative than AUROC for heavily imbalanced data~\citep{saito2015}. We report mean $\pm$ standard deviation across folds.

\subsection{Feature Ablation Analysis}

Seven configurations were evaluated: (1)~all features, (2)~no literature, (3)~literature only, (4)~genetics only, (5)~genetics plus literature, (6)~no genetics, and (7)~no somatic mutation. Each was assessed under both cross-validation and the temporal split described below, reporting AUROC, AUPRC (mean $\pm$ SD for CV), baseline AUPRC (positive-class prevalence), absolute gain ($\Delta$AUPRC), and the AUPRC-to-baseline ratio.

\subsection{SHAP Interpretability}

SHapley Additive exPlanations~\citep{lundberg2017} were computed using \texttt{TreeExplainer} in interventional mode (\texttt{feature\_perturbation='interventional'}, \texttt{model\_output='raw'}) on an XGBoost model trained on the full dataset. Mean absolute SHAP values ranked feature importance; beeswarm plots showed direction and spread. SHAP values reflect feature reliance for class separation and do not imply causation.

\subsection{Feature Correlation and Multicollinearity Diagnostics}

Spearman rank correlation matrices were computed globally and within oncology. Variance inflation factors were calculated to check for multicollinearity. SHAP dependence plots visualised the conditional relationship between somatic mutation SHAP values and literature mining scores.

\subsection{Genetic Enrichment Testing}

The association between genetic evidence (genetic association score $> 0$) and approval was tested globally and per therapeutic area using Fisher's exact test. Odds ratios are reported with 95\% confidence intervals (Woolf's log method). Benjamini--Hochberg FDR correction was applied across the six therapeutic area comparisons; Bonferroni-corrected values are given for reference. Because a single target gene can appear in multiple disease pairs, we additionally performed a target-level analysis: for each unique Ensembl ID we recorded whether any associated pair carried genetic evidence and whether any reached approval, then recomputed the OR with 2,000-iteration bootstrap confidence intervals. This target-level analysis is the primary sensitivity check for non-independence.

\subsection{Temporal Validation}
\label{sec:methods_temporal}

The training set contained all pairs whose earliest associated drug approval fell on or before 2015, plus all Phase~1/2 pairs not allocated to the test set ($n_{\text{train}} = 23{,}403$; 1,189 approved). The test set contained pairs whose earliest approval was after 2015 ($n = 150$), plus 2,700 Phase~1/2 pairs sampled to preserve the training-set class ratio ($n_{\text{test}} = 2{,}850$). The genetic enrichment hypothesis was tested independently on this temporal test set.

This design tests one important aspect of potential leakage: whether the genetic enrichment signal persists when training and test approvals are temporally separated. It does not fully eliminate leakage from literature mining, because Open Targets literature scores carry no timestamps. A pair approved in 2016 may have literature scores reflecting publications accumulated through 2024, and training-set features are similarly not frozen at the 2015 cutoff. A fully prospective test would require time-stamped evidence snapshots that Open Targets does not currently provide.

\subsection{Genetic Evidence Threshold Sensitivity}

Fisher's exact tests were repeated at thresholds of $> 0.5$ and $> 0.8$ (Supplementary Table~S2). The genetic association feature was also binarised at these thresholds inside the full XGBoost model to confirm that the continuous encoding is appropriate.

\subsection{Software and Reproducibility}

Analyses were run in Python~3.10.0 with scikit-learn~1.3.2, XGBoost~1.7.6~\citep{chen2016}, SHAP~0.42.1~\citep{lundberg2017}, pandas~2.0.3, NumPy~1.24.3, and SciPy~1.10.1. Code, a pinned dependency file (\texttt{requirements.txt}), and the curated dataset are available at \url{https://github.com/vi-c-ky/Human-genetic-evidence-associated-with-drug-approval}. Only publicly available data were used; no ethics approval was required.

\section{Results}

\subsection{Dataset Characteristics}

The final dataset comprised 26,278 target--disease pairs spanning 1,434 unique targets and 1,562 unique diseases: 1,364 (5.2\%) approved at Phase~4, 24,914 (94.8\%) at Phase~1/2. Oncology was the largest subset ($n = 10{,}699$, 658 approved), followed by Other ($n = 12{,}813$), CVRM ($n = 1{,}345$), Immunology ($n = 877$), Respiratory ($n = 370$), and Rare Disease ($n = 174$). Literature mining had non-zero scores for 53.3\% of pairs; genetic association was present in only 6.3\% (Supplementary Table~S6).

\subsection{Genetic Evidence Enriched Among Approved Targets}
\label{sec:enrichment}

\begin{figure}[H]
\centering
\includegraphics[width=0.85\textwidth]{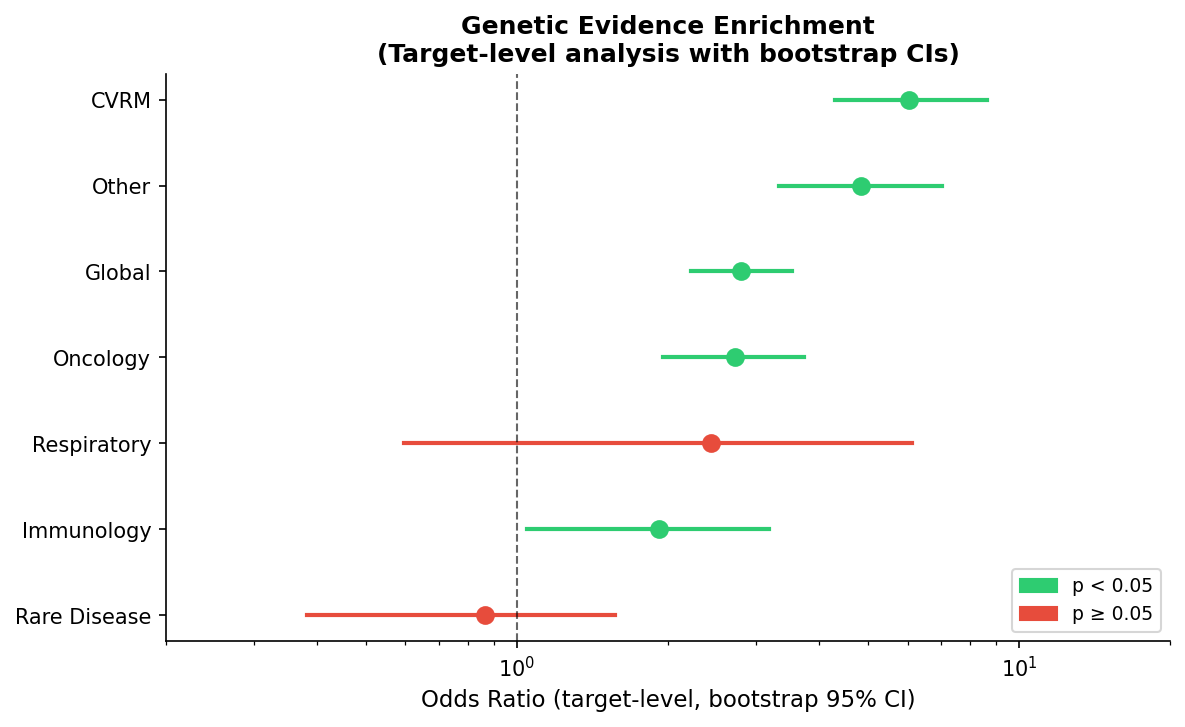}
\caption{Target-level genetic enrichment odds ratios by therapeutic area (one observation per gene, bootstrap 95\% CIs, $n_{\text{boot}} = 2{,}000$). Green: $p < 0.05$; red: $p \geq 0.05$. Dashed line at OR~$= 1$.}
\label{fig:enrichment}
\end{figure}

Globally, targets with genetic evidence had an approval rate of 13.6\% versus 4.6\% for those without, a 3.25-fold enrichment (OR~$= 3.25$, 95\% CI 2.79--3.79, $p = 1.91 \times 10^{-42}$; Figure~\ref{fig:enrichment}). At stricter thresholds (score $> 0.5$ and $> 0.8$), odds ratios rose to 4.47 (95\% CI 3.64--5.48) and 5.06 (95\% CI 3.61--7.09), suggesting a dose--response pattern (Supplementary Table~S2). These results extend the observations of Nelson et al.~\citep{nelson2015} to a contemporary multi-evidence dataset. As in their work, the observational design cannot establish a causal role for genetic evidence.

Because the 26,278 pairs derive from only 1,434 unique target genes, pair-level analyses treat correlated observations independently---a gene such as \textit{EGFR} contributes dozens of disease pairs, each counted separately. A target-level analysis (collapsing to one observation per gene) yielded OR~$= 2.79$ (bootstrap 95\% CI 2.22--3.53, $n = 1{,}434$ targets). The attenuation from 3.25 to 2.79 is expected when correlated observations are removed; the enrichment remains strong.

Across therapeutic areas (Table~\ref{tab:enrichment}), pair-level and target-level ORs can differ substantially when targets vary in the number of disease pairs they contribute. In oncology, heavily studied genes (\textit{EGFR}, \textit{BRAF}) each contribute many correlated pairs, so the pair-level OR (6.72) inflates relative to the target-level OR (2.71). Conversely, in CVRM---where targets tend to have fewer disease pairs---the target-level OR (6.04, bootstrap 95\% CI 4.31--8.66) exceeds the pair-level estimate (2.64). Target-level estimates are therefore the more reliable summary for cross-area comparisons. Immunology reached significance at the target level (OR~$= 1.92$, 95\% CI 1.05--3.18). Respiratory trended positive but did not reach significance (OR~$= 2.44$, 95\% CI 0.60--6.13, $p = 0.08$), and rare disease showed no enrichment (OR~$= 0.86$, 95\% CI 0.38--1.56). All pair-level $p$-values in Table~\ref{tab:enrichment} are Benjamini--Hochberg adjusted across six comparisons.

The ``Other'' category warrants particular caution: the pair-level OR of 1.50 and target-level OR of 4.84 reflect a heterogeneous mixture of disease areas with markedly different approval rates and genetic architectures. The subcategory breakdown is described in Section~\ref{sec:other}.

\begin{table}[H]
\centering
\caption{Genetic evidence enrichment by therapeutic area. $p$-values are Benjamini--Hochberg adjusted across six comparisons.}
\label{tab:enrichment}
\small
\begin{tabular}{lccccc}
\toprule
Therapeutic Area & OR & 95\% CI & $p_{\text{raw}}$ & $p_{\text{adj}}$ (BH) \\
\midrule
Oncology & 6.72 & 5.34--8.47 & $1.22 \times 10^{-48}$ & $7.34 \times 10^{-48}$ \\
Respiratory & 4.33 & 1.75--10.72 & $5.50 \times 10^{-3}$ & $1.10 \times 10^{-2}$ \\
CVRM & 2.64 & 1.81--3.85 & $3.76 \times 10^{-6}$ & $1.13 \times 10^{-5}$ \\
Other & 1.50 & 1.05--2.16 & $3.17 \times 10^{-2}$ & $4.75 \times 10^{-2}$ \\
Immunology & 1.76 & 0.95--3.24 & 0.078 & 0.093 \\
Rare Disease & 1.15 & 0.46--2.86 & 0.802 & 0.802 \\
\bottomrule
\end{tabular}
\end{table}

\subsection{Model Performance}

Under target-grouped cross-validation, both classifiers produced modest but above-baseline discrimination (Table~\ref{tab:performance}). Logistic regression (AUROC~$= 0.638 \pm 0.012$, AUPRC~$= 0.121 \pm 0.015$) slightly outperformed XGBoost (AUROC~$= 0.619 \pm 0.016$, AUPRC~$= 0.109 \pm 0.017$) against a baseline AUPRC of 0.052. Standard pair-level CV gave near-identical results (XGBoost AUROC~$= 0.622 \pm 0.018$, AUPRC~$= 0.109 \pm 0.010$), confirming that target-level leakage does not meaningfully inflate performance estimates. The near-identical performance of linear and non-linear models suggests a predominantly additive signal, consistent with the feature sparsity and literature-mining dominance described in Section~\ref{sec:ablation}. In absolute terms, the best model gains 6.9 percentage points of AUPRC over baseline. The model is poorly calibrated: predicted probabilities range from 0.25 to 0.75, but observed positive rates never exceed 12\% (Supplementary Figure~S5).

\begin{table}[H]
\centering
\caption{Cross-validated model performance. Primary results use 5-fold \texttt{StratifiedGroupKFold} (grouped by target gene). Standard pair-level CV shown for comparison.}
\label{tab:performance}
\begin{tabular}{lcccc}
\toprule
Model & CV Type & AUROC & AUPRC & AUPRC / Baseline \\
\midrule
Logistic Regression & Grouped & $0.638 \pm 0.012$ & $0.121 \pm 0.015$ & 2.33$\times$ \\
XGBoost & Grouped & $0.619 \pm 0.016$ & $0.109 \pm 0.017$ & 2.10$\times$ \\
XGBoost & Pair-level & $0.622 \pm 0.018$ & $0.109 \pm 0.010$ & 2.10$\times$ \\
Random Baseline & --- & 0.500 & 0.052 & 1.00$\times$ \\
\bottomrule
\end{tabular}
\end{table}

\subsection{Feature Ablation Analysis}
\label{sec:ablation}

Ablation results are summarised in Table~\ref{tab:ablation}.

Literature mining alone reaches near-full-model performance (grouped CV AUPRC~$= 0.099 \pm 0.022$, $1.91\times$ baseline; temporal AUPRC~$= 0.120$, $2.27\times$). That a single feature captures most discriminative ability is consistent with temporal leakage---approved targets accumulate publications after approval, inflating their literature scores---though we cannot rule out that some fraction reflects genuine pre-approval research activity.

Dropping literature leaves the remaining five evidence types above baseline (grouped CV AUPRC~$= 0.084 \pm 0.008$, $1.63\times$; temporal AUPRC~$= 0.072$, $1.37\times$), confirming that non-literature evidence carries real information. Genetic evidence in isolation provides only modest discrimination (grouped CV AUPRC~$= 0.062 \pm 0.003$, $1.19\times$). With genetic scores non-zero in just 6.3\% of pairs, the enrichment signal is better captured by the binary Fisher's exact test (Section~\ref{sec:enrichment}) than by a classifier operating on a sparse continuous score.

Removing genetics from the full model barely changes performance (grouped CV AUPRC $0.109 \to 0.106$), indicating partial redundancy with other features. Removing somatic mutation nudges grouped CV AUPRC down slightly ($0.109 \to 0.107$) but improves temporal AUPRC markedly ($0.099 \to 0.147$). The interpretation of this pattern is addressed in Section~\ref{sec:somatic}.

\begin{table}[H]
\centering
\caption{Feature ablation. XGBoost under systematic feature removal. CV uses \texttt{StratifiedGroupKFold} (grouped by target gene); values are mean $\pm$ SD across five folds. Baseline AUPRC equals positive-class prevalence. $\Delta$AUPRC is the absolute gain. $^{\dagger}$Literature scores are not time-stamped; temporal AUPRC for configurations including literature may be inflated by post-approval publications.}
\label{tab:ablation}
\small
\begin{tabular}{lcccccccc}
\toprule
 & \multicolumn{4}{c}{Grouped cross-validation} & \multicolumn{4}{c}{Temporal split} \\
\cmidrule(lr){2-5} \cmidrule(lr){6-9}
Configuration & AUPRC & $\Delta$AUPRC & Baseline & Ratio & AUROC & AUPRC & Baseline & Ratio \\
\midrule
All features & $.109 \pm .017$ & +.057 & .052 & 2.10$\times$ & .596 & .099 & .053 & 1.89$\times$ \\
No literature & $.084 \pm .008$ & +.033 & .052 & 1.63$\times$ & .532 & .072 & .053 & 1.37$\times$ \\
Literature only & $.099 \pm .022$ & +.047 & .052 & 1.91$\times$ & .626 & .120$^{\dagger}$ & .053 & 2.27$\times$ \\
Genetics only & $.062 \pm .003$ & +.010 & .052 & 1.19$\times$ & .483 & .064 & .053 & 1.22$\times$ \\
Genetics + lit. & $.092 \pm .014$ & +.040 & .052 & 1.77$\times$ & .633 & .128 & .053 & 2.44$\times$ \\
No genetics & $.106 \pm .017$ & +.055 & .052 & 2.05$\times$ & .599 & .104 & .053 & 1.98$\times$ \\
No somatic & $.107 \pm .018$ & +.055 & .052 & 2.06$\times$ & .619 & .147 & .053 & 2.79$\times$ \\
\bottomrule
\end{tabular}
\end{table}

\textbf{Threshold sensitivity.} Binarising the genetic association feature at 0, 0.5, and 0.8 inside the full XGBoost model yielded CV AUPRC of 0.109, 0.109, and 0.108. Performance is insensitive to threshold, confirming that the continuous score adds no discriminative information beyond the binary flag.

\subsection{Temporal Validation}

Training on pre-2015 approvals and testing on post-2015 approvals gave AUROC~$= 0.596$ and AUPRC~$= 0.099$ ($1.89\times$ baseline), a modest drop from cross-validation consistent with reducing temporal leakage (Table~\ref{tab:ablation}). Excluding literature, above-baseline performance persisted (AUROC~$= 0.532$, AUPRC~$= 0.072$, $1.37\times$).

The genetic enrichment finding replicated in the temporal test set (OR~$= 3.51$, $p = 1.72 \times 10^{-8}$). Because the binary genetic flag is unlikely to change materially after approval, this replication is credible. As noted in Section~\ref{sec:methods_temporal}, however, the temporal design does not fully remove literature-mining leakage, and evidence scores for both training and test pairs reflect data accumulated through 2024.

\subsection{SHAP Feature Importance}

\begin{figure}[H]
\centering
\includegraphics[width=0.85\textwidth]{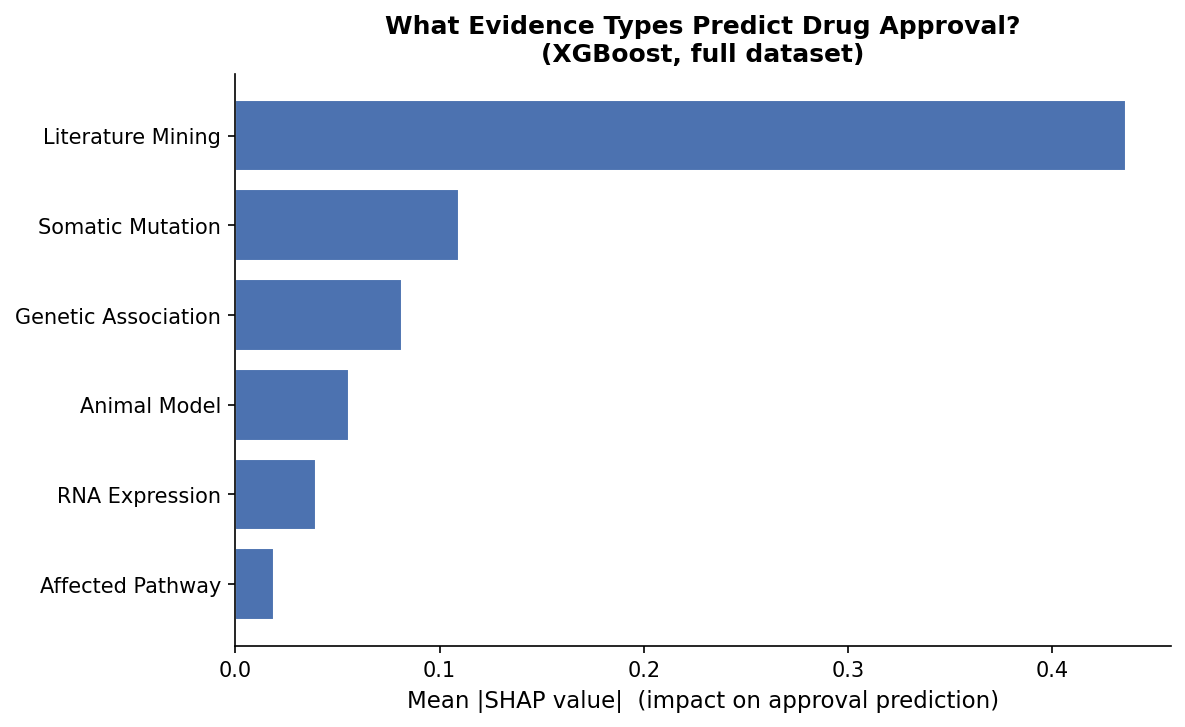}

\medskip
\includegraphics[width=0.9\textwidth]{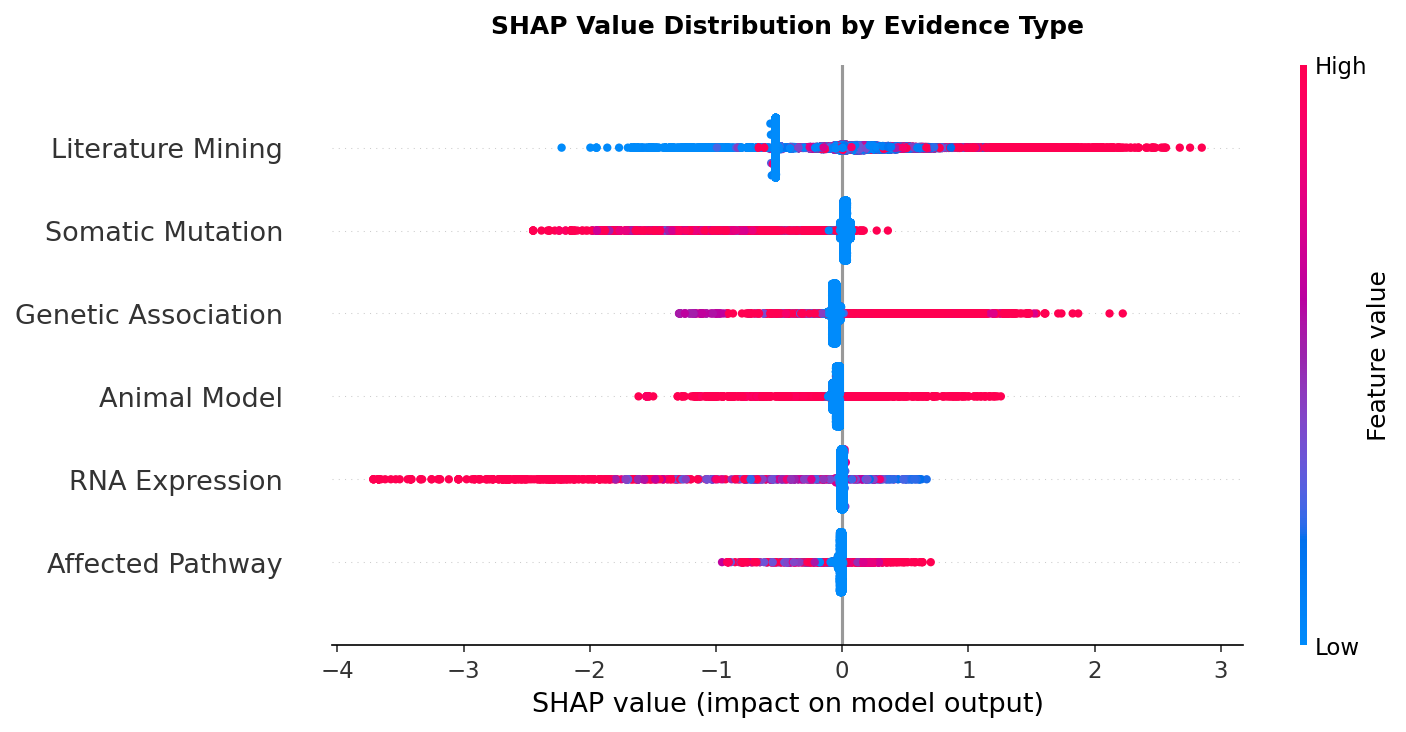}
\caption{SHAP analysis of the XGBoost model. (a)~Mean absolute SHAP values showing literature mining dominance (mean $|\text{SHAP}| = 0.44$), likely reflecting temporal leakage. (b)~Beeswarm plot: high literature and genetic scores increase predicted approval; high somatic mutation scores decrease it conditional on other features.}
\label{fig:shap_combined}
\end{figure}

The model relies most heavily on literature mining (mean $|\text{SHAP}| = 0.44$), with somatic mutation (0.11) and genetic association (0.08) distant (Figure~\ref{fig:shap_combined}a). Given the ablation results, this ranking most likely reflects leakage rather than genuine predictive primacy. The beeswarm plot (Figure~\ref{fig:shap_combined}b) shows that high literature and genetic scores push predicted approval upward, while high somatic mutation scores pull it downward once other features are accounted for (Section~\ref{sec:somatic}).

\subsection{Somatic Mutation: Suppression in the Multivariate Model}
\label{sec:somatic}

Somatic mutation has a positive univariate association with approval (OR~$= 1.73$, 95\% CI 1.48--2.02, $p = 5.17 \times 10^{-11}$), yet its SHAP values in the multivariate model are negative. This apparent reversal is a statistical suppression effect: somatic mutation and literature mining share moderate variance ($\rho = 0.36$ globally, $0.38$ in oncology; Supplementary Table~S7), and once the shared variance with literature is partialled out, the remaining somatic mutation signal is negatively associated with approval in the model.

VIF values were low everywhere (global 1.02--1.53; oncology 1.01--1.64), ruling out numerical instability as an explanation. The pattern is confirmed by stratified approval rates: among somatic-mutation-positive pairs, approval rose from 1.5\% in the lowest literature quartile to 18.8\% in the highest, indicating that approval is predominantly associated with co-occurring literature rather than somatic mutation per se.

The suppression is statistical in origin. Whether the residual negative signal also reflects enrichment for clinically intractable oncogenic drivers~\citep{dang2017} would require gene-level analysis beyond our scope.

\subsection{Candidate Opportunity Targets}

\begin{figure}[H]
\centering
\includegraphics[width=\textwidth]{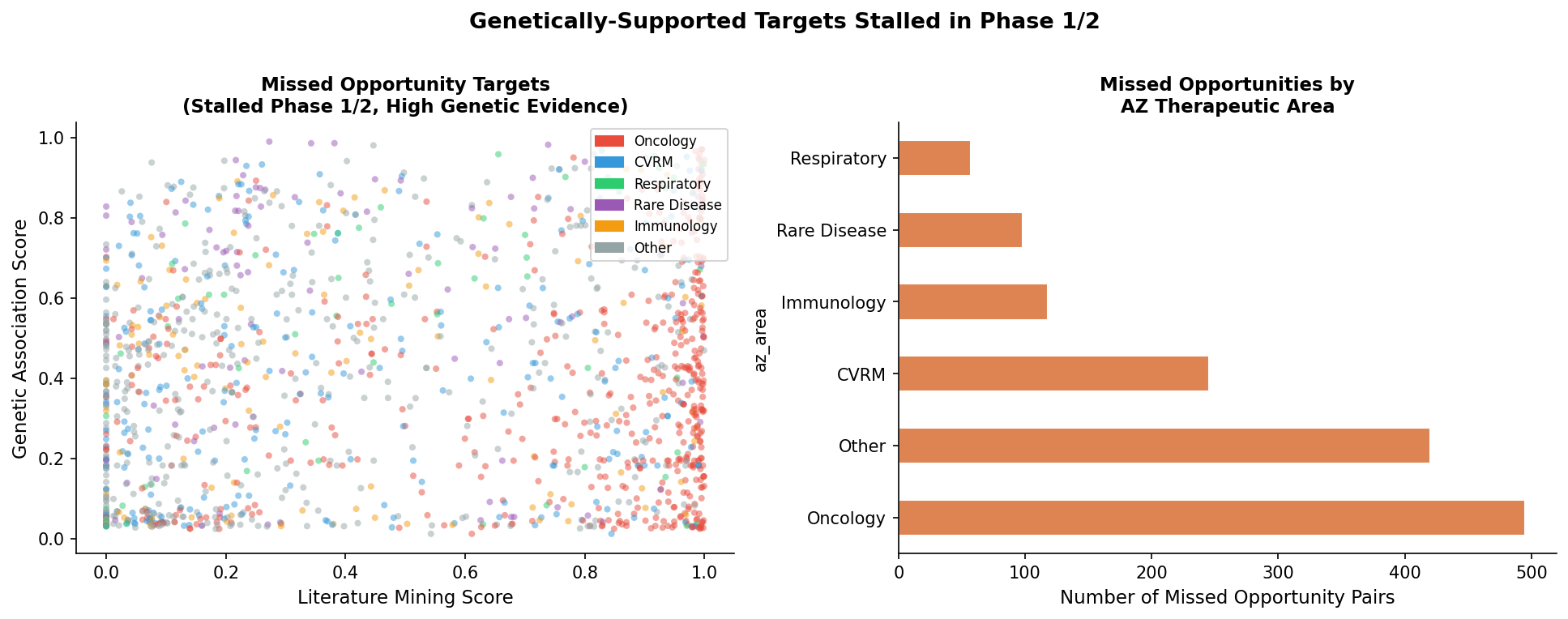}
\caption{Genetically supported target--disease pairs at Phase~1/2 ($n = 1{,}433$). Left: scatter by genetic and literature scores. Right: distribution by therapeutic area.}
\label{fig:missed}
\end{figure}

A total of 1,433 target--disease pairs at Phase~1/2 carry non-zero genetic association scores: 495 (34.5\%) in oncology, 420 (29.3\%) in ``Other,'' 245 (17.1\%) in CVRM, 118 (8.2\%) in immunology, 98 (6.8\%) in rare disease, and 57 (4.0\%) in respiratory (Figure~\ref{fig:missed}).

Plotting genetic versus literature scores reveals two clusters: one with high genetic but low literature scores (genetically implicated targets with little publication activity) and one with high scores on both axes (extensively studied targets that remain unapproved, possibly for druggability or therapeutic-index reasons rather than target-selection failures). Targets in the high-genetic/low-literature cluster---with strong genetic support but low literature scores---are biologically implicated by human genetic evidence but have attracted little experimental follow-up, and represent the more defensible ``under-studied'' subgroup within this catalogue.

``Phase~1/2 status'' does not equate to failure. Some pairs may be in active development; others may have been shelved for commercial reasons. The list is a hypothesis-generating starting point, not a ranked prioritisation. The full list is in Supplementary Table~S1.

\subsection{Composition of the ``Other'' Therapeutic Area}
\label{sec:other}

The heterogeneous composition of the ``Other'' bin motivated a pre-specified subgroup analysis of its five largest disease subcategories to characterise the sources of its aggregate enrichment signal. The ``Other'' bin ($n = 7{,}001$) is heterogeneous. Its five largest subcategories are phenotype/trait associations ($n = 1{,}624$, 6.0\% approved), psychiatric disorders ($n = 1{,}359$, 12.1\%), nervous system diseases ($n = 892$, 4.4\%), infectious diseases ($n = 526$, 1.3\%), and urinary system diseases ($n = 461$, 0.9\%). Genetic enrichment was significant for phenotype associations (OR~$= 3.53$, $p = 2.86 \times 10^{-3}$) and nervous system diseases (OR~$= 5.79$, $p = 1.81 \times 10^{-4}$), both surviving Bonferroni correction for five subcategory tests. Psychiatric disorders showed no enrichment (OR~$= 0.68$, $p = 0.30$). The heterogeneity of this category means that its aggregate pair-level OR of 1.50 should not be interpreted as a single biological signal. The null finding in psychiatric disorders---the second-largest subcategory---warrants follow-up given the clinical relevance of CNS targets and the known challenges of translating polygenic GWAS signals to therapies.

\subsection{Sensitivity Analyses}
\label{sec:sensitivity}

\textbf{Phase~3 inclusion.} Adding the 12,125 excluded Phase~3 pairs as negatives ($n = 38{,}362$) strengthened enrichment. The global OR rose from 3.25 to 4.93 ($p = 2.18 \times 10^{-70}$), and all per-area ORs increased (e.g., oncology 6.72 $\to$ 9.01; CVRM 2.64 $\to$ 4.01). Immunology reached significance (OR~$= 2.17$, $p = 1.78 \times 10^{-2}$). Phase~3 pairs without genetic support expand the negative class, so this increase is expected; the primary analysis excluding Phase~3 is therefore conservative.

\textbf{Likely-terminated negatives.} Restricting to the 16,693 pairs where the associated drug was approved for another indication before 2015 gave OR~$= 4.54$ ($p = 1.15 \times 10^{-60}$), higher than the primary estimate. This reinforces the view that contamination of the negative class by ongoing programmes biases the primary OR toward the null.

\section{Discussion}
\label{sec:discussion}

The central finding of this analysis is that genetic evidence is robustly enriched among approved drug targets. The pair-level OR of 3.25 (95\% CI 2.79--3.79) attenuates to 2.79 (bootstrap CI 2.22--3.53) at the target level, confirming the association after accounting for non-independence. The enrichment replicates temporally (OR~$= 3.51$) and strengthens at stricter genetic score thresholds. Sensitivity analyses bracket the plausible range between 3.25 and 4.93, with all estimates substantially above the null. The direction, magnitude, and consistency of this association across analyses is the main contribution of the study.

The therapeutic-area results reveal meaningful heterogeneity. At the target level---the more reliable summary for areas with variable numbers of pairs per gene---CVRM shows the strongest enrichment (OR~$= 6.04$), while rare disease shows none (OR~$= 0.86$). The null result in rare disease most likely reflects the fact that genetic evidence (often from Mendelian disease genetics) already guides target selection in that area, making genetically supported and unsupported targets non-comparable populations. Psychiatric disorders within the ``Other'' category also showed no enrichment; polygenic architectures and blood--brain barrier penetration challenges may limit the translational utility of GWAS-derived targets in this area, though this subgroup analysis is exploratory. These area-specific patterns merit follow-up with larger, better-annotated datasets.

Literature mining accounts for most classifier performance, consistent with temporal leakage. Literature alone achieves CV AUPRC~$= 0.099$, close to the full-model value of 0.109; in the temporal split, the literature-only AUPRC (0.120) nominally exceeds the full model (0.099). This pattern is consistent with approved targets accumulating publications after regulatory approval. However, because Open Targets literature scores carry no timestamps, we cannot fully separate pre-approval research activity from post-approval accumulation. The SHAP ranking placing literature above all other features should therefore be understood as reflecting a data composition artefact rather than a true predictive hierarchy. The ablation---comparing performance with and without literature---bounds the leakage contribution but does not eliminate it.

The machine-learning models have limited practical predictive value. The best model gains 6.9 percentage points of AUPRC over baseline, predicted probabilities are poorly calibrated relative to observed approval rates, and the genetics-only configuration contributes only a 1.0-percentage-point absolute gain. The genetic enrichment signal is better characterised as binary---whether genetic evidence is present or absent---than as a continuous score grading approval probability. The comparable performance of logistic regression and XGBoost further suggests that any signal in this dataset is predominantly additive rather than reflecting complex feature interactions.

\subsection*{Limitations}

\textbf{Negative-class heterogeneity} is the most important caveat. Phase~1/2 pairs include ongoing programmes and commercially abandoned projects alongside true failures. This contamination biases odds ratios toward the null, so the primary estimates are conservative. The sensitivity analyses (Section~\ref{sec:sensitivity}) bracket the plausible range but cannot resolve which individual pairs are truly terminated.

\textbf{Literature-mining leakage} cannot be fully characterised with the available data. The temporal split credibly tests the genetic enrichment finding but does not remove leakage from literature mining, whose scores are not timestamped. Disentangling genuine pre-approval research activity from post-approval accumulation would require publication-date-aware features that Open Targets does not currently provide.

\textbf{Non-independence of target--disease pairs} is addressed by both target-grouped cross-validation and target-level OR analyses. The near-identical grouped and pair-level CV metrics confirm that target-level leakage does not inflate performance estimates. Target-level ORs, however, differ substantially from pair-level ORs in areas with highly multiplexed targets (oncology pair-level OR 6.72 vs target-level 2.71), and the target-level estimates are the more reliable cross-area summary.

\textbf{Confounding by druggability and target-selection practice.} Genetically validated targets may be preferentially pursued because of their genetic evidence, creating a selection effect that the present observational design cannot disentangle from any direct contribution of genetic support to approval probability. This confounding applies to all evidence types.

\textbf{Heterogeneity of the ``Other'' category.} Nearly half the dataset falls into this bin, which aggregates disease areas with markedly different approval rates and genetic architectures. Its aggregate OR should not be interpreted as a single biological signal.

\textbf{ChEMBL coverage.} Coverage of early-phase failures is incomplete. The dataset captures drugs with both mechanism and indication records, which may over-represent commercially active programmes and under-represent shelved ones.

This analysis is correlational throughout. No causal claims about the effect of genetic evidence on approval are warranted.

\subsection*{Future Work}

Incorporating Phase~3 outcomes as they mature, adding Mendelian randomisation evidence as a distinct genetic feature, and using network-based target representations are natural extensions. Linking to clinical-trial registry data would improve failure coverage. Time-stamped evidence scores would permit cleaner temporal analyses. The strongest test of practical utility would be a prospective evaluation applied to ongoing Phase~1/2 programmes.

\section{Conclusions}

Genetic evidence is consistently enriched among approved drug targets across therapeutic areas (pair-level OR~$= 3.25$, 95\% CI 2.79--3.79; target-level OR~$= 2.79$, bootstrap 95\% CI 2.22--3.53). The association replicates in temporal validation (OR~$= 3.51$) and is robust across a range of negative-class definitions (sensitivity OR range 3.25--4.93). A target-level analysis accounting for the non-independence of pairs sharing the same gene confirms that the signal is not an artefact of repeated observations, and the magnitude of the oncology pair-level OR (6.72) is substantially attenuated at the target level (2.71), underscoring why this correction matters.

Feature ablation identifies literature mining as the dominant contributor to classifier performance, consistent with temporal leakage from post-approval publications; because literature scores carry no timestamps, this conclusion is supported but not definitive. Excluding literature, remaining evidence types retain above-baseline signal, and the genetic enrichment finding stands across all analytical specifications.

The machine-learning models have modest discrimination and poor calibration and are not suitable for practical deployment. The genetic enrichment signal is better understood as a binary population-level association than as a continuous predictor of individual approval outcomes.

We release 1,433 genetically supported Phase~1/2 pairs as a hypothesis-generating resource. These findings are observational and support genetic evidence as a consistent correlate of drug approval, while the limitations of the current data preclude stronger conclusions.

\section*{Acknowledgements}

Data were obtained from the Open Targets Platform and ChEMBL database. We thank the maintainers of these open-access resources. Claude (Anthropic) was used as a writing and analytical assistant during manuscript preparation. All scientific content, analyses, and interpretations were verified by the authors, who take full responsibility for accuracy.

\section*{Data Availability}

Open Targets Platform and ChEMBL (v33) data are publicly available via their respective APIs. The curated dataset of 26,278 target--disease pairs, the candidate list of 1,433 Phase~1/2 pairs, and all analysis code are available at \url{https://github.com/vi-c-ky/Human-genetic-evidence-associated-with-drug-approval}.

\section*{Supplementary Information}

\textbf{Supplementary Table~S1.} Full list of 1,433 genetically supported target--disease pairs (CSV).

\textbf{Supplementary Table~S2.} Genetic enrichment odds ratios at higher score thresholds (global and by therapeutic area), with 95\% CIs.

\textbf{Supplementary Table~S3.} Sensitivity analyses: Phase~3 inclusion and likely-terminated negative class restriction.

\textbf{Supplementary Table~S4.} ``Other'' therapeutic area breakdown by disease subcategory.

\textbf{Supplementary Table~S5.} Binarised genetic feature ablation results.

\textbf{Supplementary Table~S6.} Evidence type coverage in the final dataset.

\textbf{Supplementary Table~S7.} Spearman rank correlations between evidence types (global and oncology).

\textbf{Supplementary Figure~S1.} ROC and Precision--Recall curves for logistic regression and XGBoost.

\textbf{Supplementary Figure~S2.} Spearman correlation matrices (global and oncology).

\textbf{Supplementary Figure~S3.} AUROC and approval rates by therapeutic area.

\textbf{Supplementary Figure~S4.} SHAP dependence plots for somatic mutation.

\textbf{Supplementary Figure~S5.} Calibration plot (reliability diagram) and predicted probability distribution for XGBoost under target-grouped cross-validation. The model is poorly calibrated: predicted probabilities cluster around 0.3--0.7 while observed positive rates range from 3--12\%, indicating the classifier assigns unrealistically high approval probabilities.

\bibliographystyle{plainnat}
\bibliography{references}

\clearpage

\begin{figure}[H]
\centering
\includegraphics[width=\textwidth]{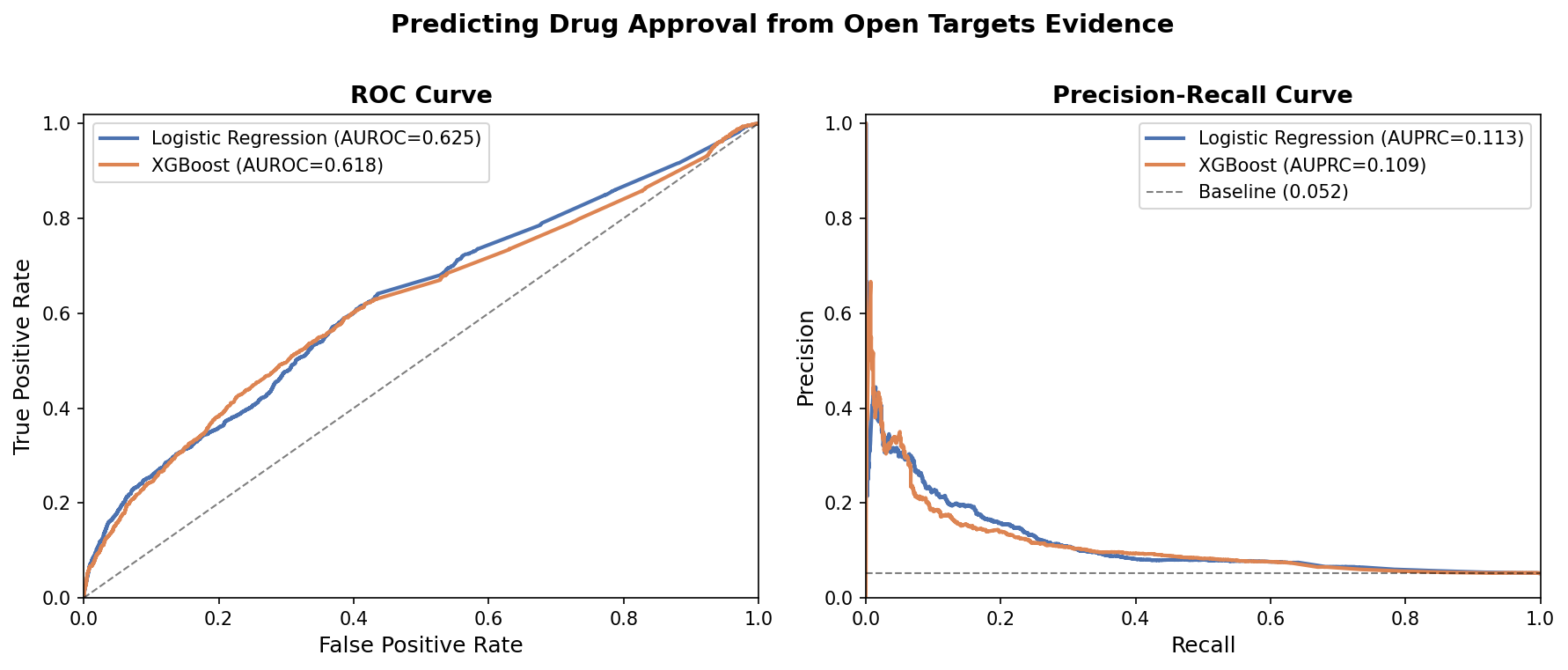}
\caption*{\textbf{Supplementary Figure S1.} ROC and Precision--Recall curves comparing Logistic Regression and XGBoost classifiers.}
\end{figure}

\begin{figure}[H]
\centering
\includegraphics[width=0.9\textwidth]{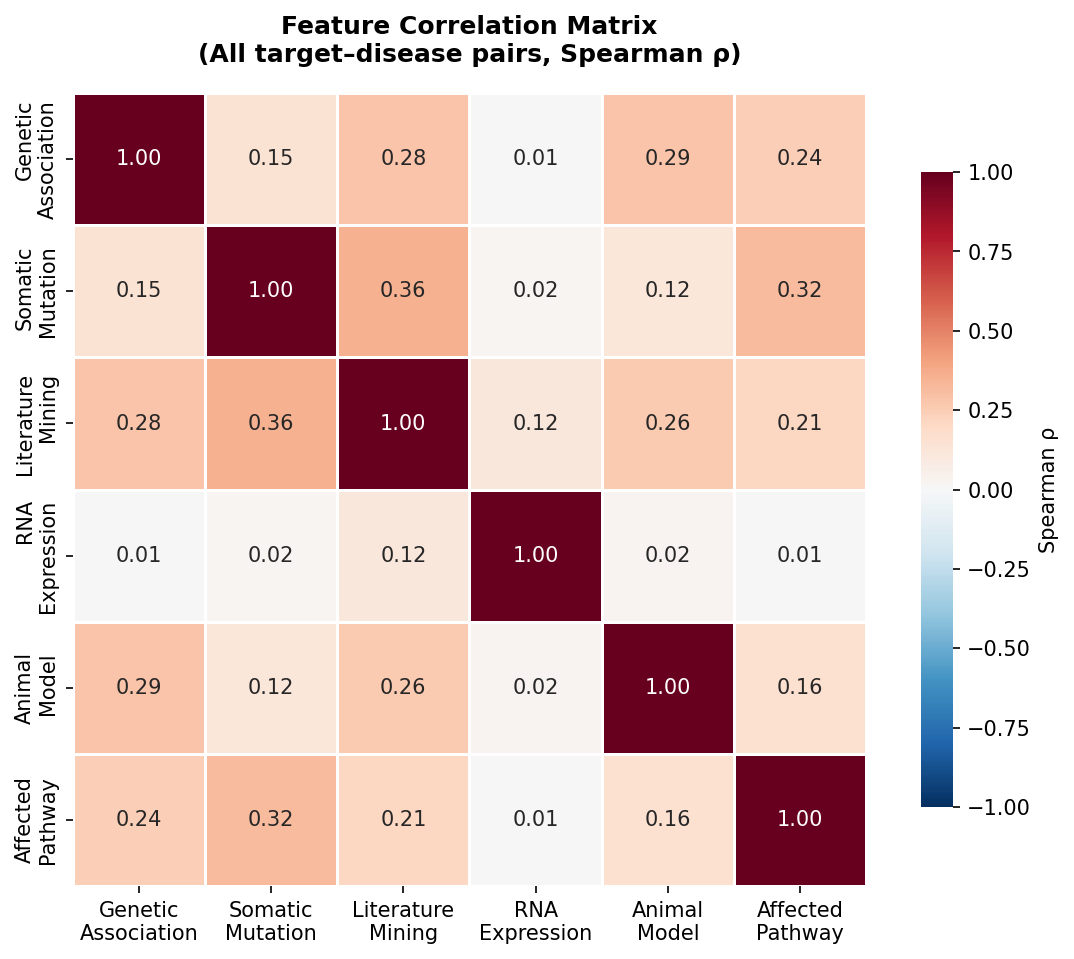}
\caption*{\textbf{Supplementary Figure S2a.} Spearman rank correlation matrix across all target--disease pairs.}
\end{figure}

\begin{figure}[H]
\centering
\includegraphics[width=0.9\textwidth]{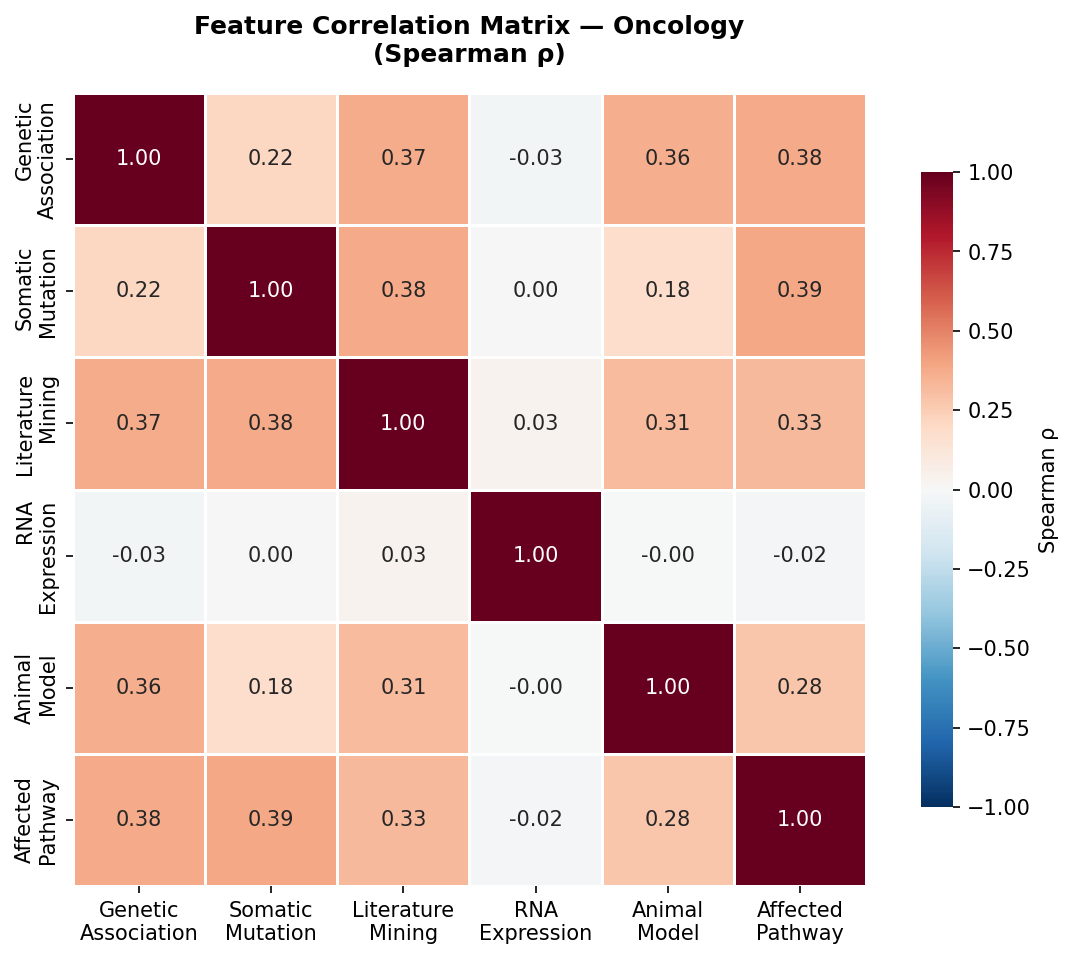}
\caption*{\textbf{Supplementary Figure S2b.} Spearman rank correlation matrix for oncology target--disease pairs.}
\end{figure}

\begin{figure}[H]
\centering
\includegraphics[width=0.9\textwidth]{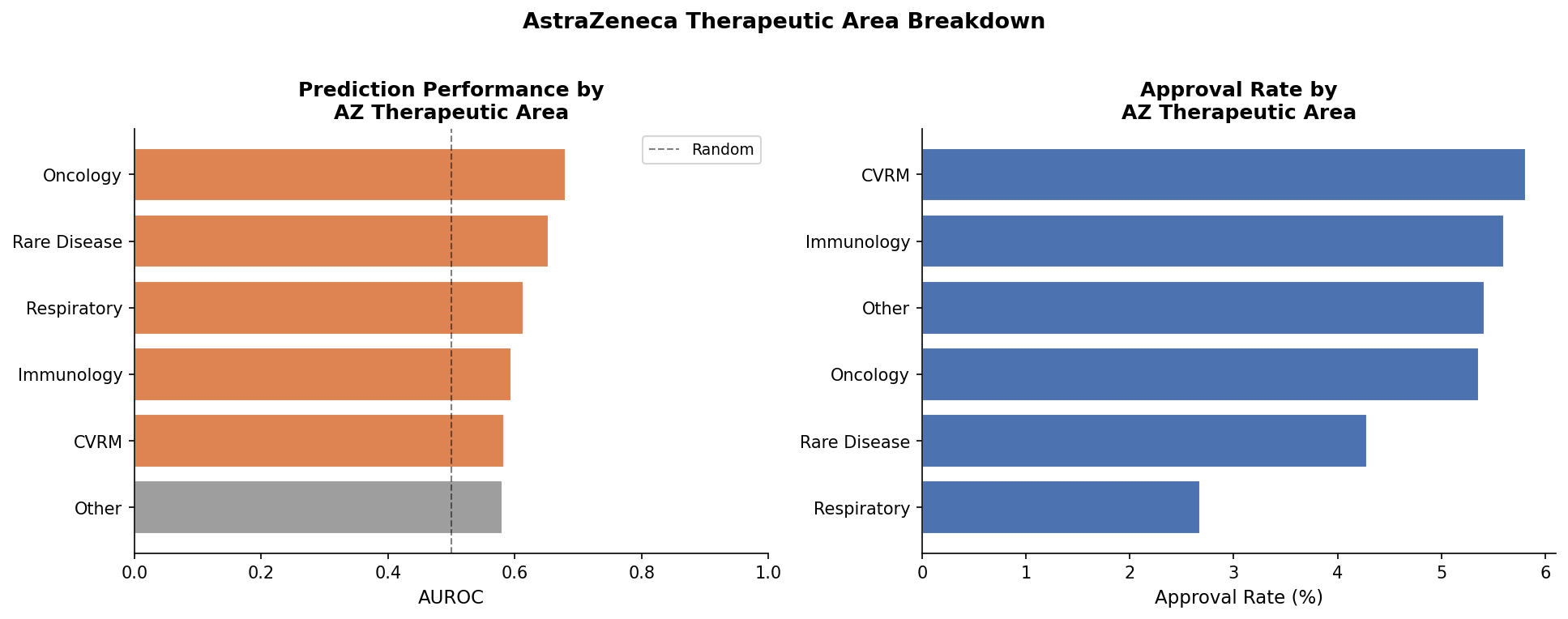}
\caption*{\textbf{Supplementary Figure S3.} Prediction performance (AUROC) and approval rates by therapeutic area.}
\end{figure}

\begin{figure}[H]
\centering
\includegraphics[width=\textwidth]{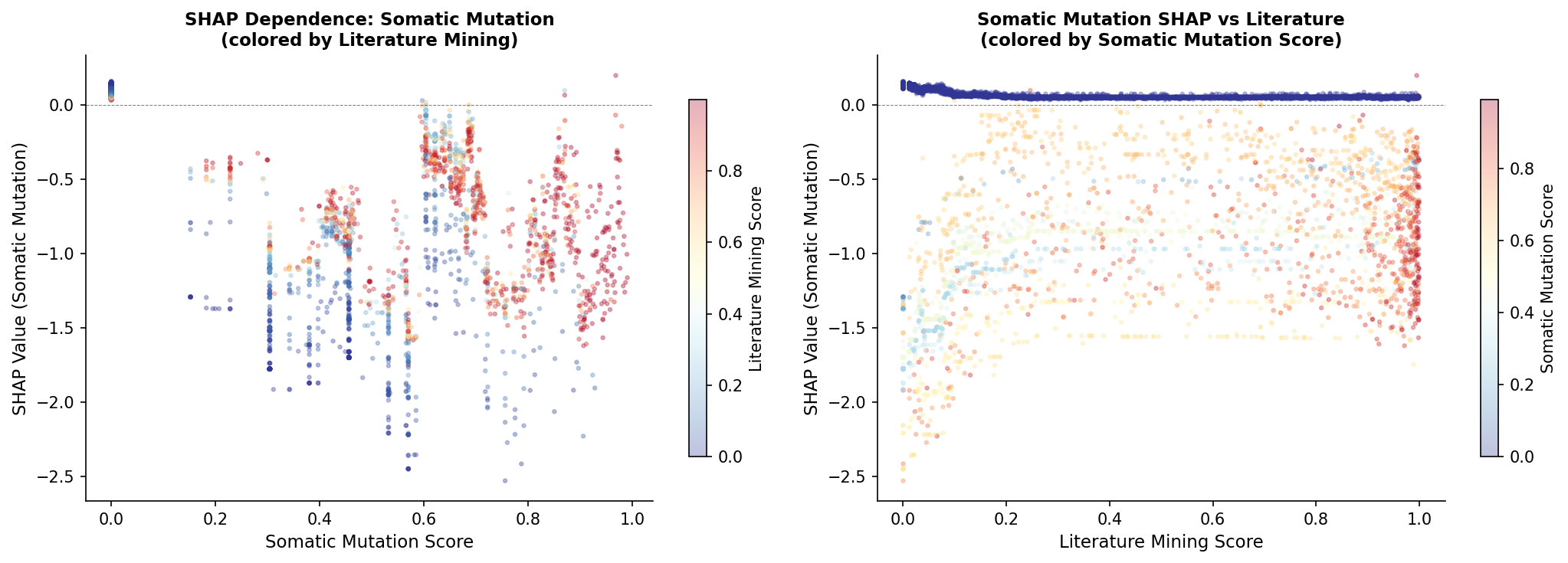}
\caption*{\textbf{Supplementary Figure S4.} SHAP dependence plots for somatic mutation. Negative SHAP values persist across literature score strata.}
\end{figure}

\begin{figure}[H]
\centering
\includegraphics[width=\textwidth]{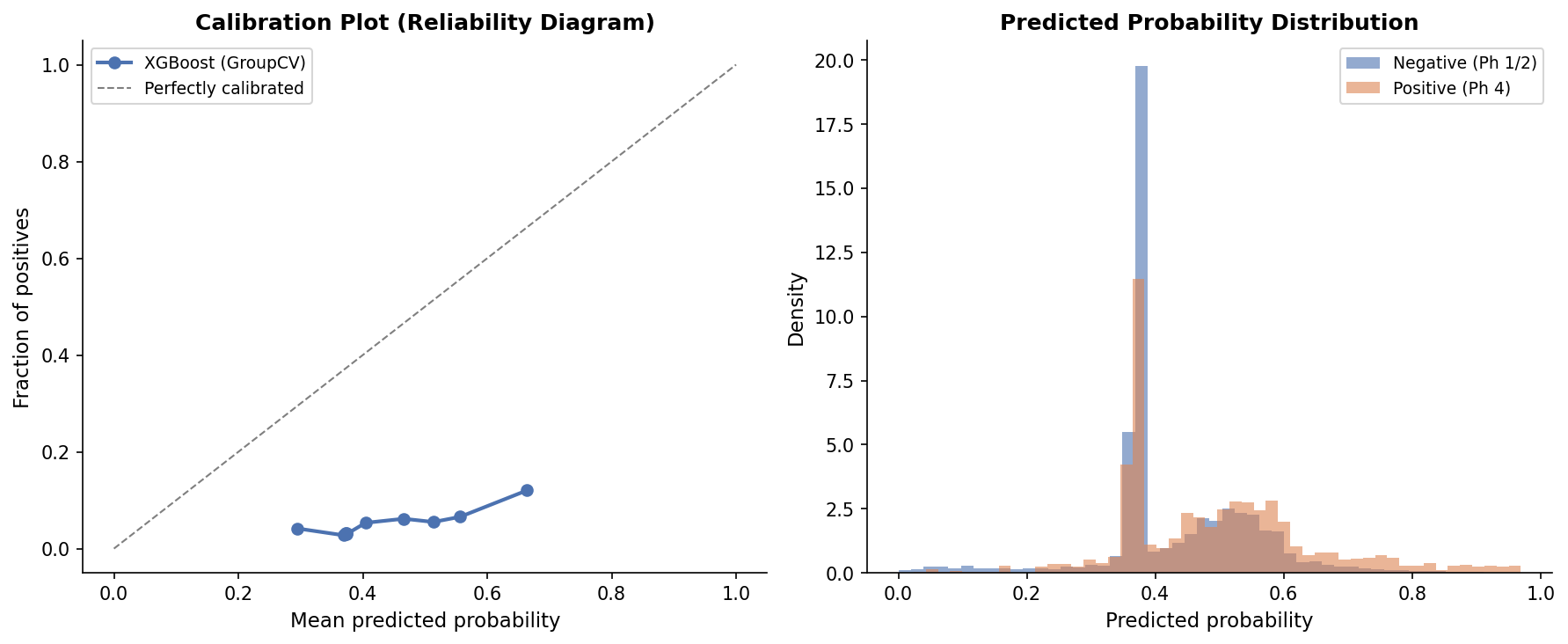}
\caption*{\textbf{Supplementary Figure S5.} Calibration plot (left) and predicted probability distribution (right) for XGBoost under target-grouped cross-validation. The model is poorly calibrated: predicted probabilities cluster around 0.3--0.7 while observed positive rates range from 3--12\%, indicating the classifier assigns unrealistically high approval probabilities.}
\end{figure}

\end{document}